\documentclass[a4paper,fleqn]{cas-dc}

\usepackage[numbers,sort&compress]{natbib}
\usepackage{import}
\usepackage{pgf} 
\usepackage{caption}
\usepackage{subcaption}

\def\tsc#1{\csdef{#1}{\textsc{\lowercase{#1}}\xspace}}
\tsc{WGM}
\tsc{QE}

\begin{document}

\let\WriteBookmarks\relax
\def\floatpagepagefraction{1}
\def\textpagefraction{.001}

\shorttitle{Descriptor-based 3D Microstructure Characterization and Reconstruction}
\shortauthors{P.\ Seibert, A.\ Ra{\ss}loff, M.\ Ambati et al.}  

\title [mode = title]{Descriptor-based reconstruction of three-dimensional microstructures through gradient-based optimization}

\author[1]{Paul Seibert}[orcid=0000-0002-8774-8462]

\author[1]{Alexander Ra{\ss}loff}[orcid=0000-0002-9134-4874]

\author[1]{Marreddy Ambati}[orcid=0000-0003-1086-1392]

\author[1,2,3]{Markus K\"astner}[orcid=0000-0003-3358-1545]
\ead{Markus.Kaestner@TU-Dresden.de}

\affiliation[1]{organization={Institute of Solid Mechanics, Technische Universit\"at Dresden},
            postcode={01069},
            city={Dresden},
            addressline={George-B\"ahr-Stra{\ss}e 3c},
            country={Germany}}
\affiliation[2]{organization={Dresden Center for Computational Materials Science, TU Dresden},
            city={Dresden},
            country={Germany}}
\affiliation[3]{organization={Dresden Center for Fatigue and Reliability},
            addressline={TU Dresden}, 
            city={Dresden},
            country={Germany}}

\cormark[1]
\cortext[4]{Corresponding author.}

\begin{abstract}
Microstructure reconstruction is an important cornerstone to the inverse materials design concept. 
In this work, a general algorithm is developed to reconstruct a three-dimensional microstructure from given descriptors. 
Based on two-dimensional (2D) micrographs, this reconstruction algorithm allows valuable insight through spatial visualization of the microstructure and in silico studies of structure-property linkages. 
The formulation ensures computational efficiency by casting microstructure reconstruction as a gradient-based optimization problem. 
Herein, the descriptors can be chosen freely, such as spatial correlations or Gram matrices, as long as they are differentiable with respect to the microstructure. 
Because real microstructure samples are commonly available as 2D microscopy images only, the desired descriptors for the reconstruction process are prescribed on orthogonal 2D slices. 
This adds a source of noise, which is handled in a new, superior and interpretable manner. 
The efficiency and applicability of this formulation is demonstrated by various numerical experiments. 
\end{abstract}


\begin{keywords}
    Microstructure \sep
    Reconstruction \sep
    3D characterization \sep
    Statistics \sep
    Gradient-based optimization
\end{keywords}

\maketitle

\section{Introduction}\label{sec:intro}
Accelerating materials development is a major goal in academia and industry, even recognized in politics, as innovative, tailored and optimized materials are seen as enabler for solving topical challenges regarding resource-efficiency and durability~\cite{nstc11}.
A promising approach for overcoming the rather long duration associated with the traditional innovation process is the inverse design concept. 
For this reason, considerable attention has been attracted to the investigation of process-microstructure-property linkages of materials, which, in the end, are intended to be used for an optimization of microstructures regarding favorable properties. Thanks to rising computational resources, numerical simulations can be utilized for building large databases that allow for the application of data-driven techniques for this purpose.
A central issue that needs to be addressed in this context is how to select and provide the massive amount of suitable microstructures for the simulations. An efficient and universally applicable \textit{microstructure characterization and reconstruction} (MCR) method is therefore not only helpful but vital for the implementation of this concept.
In a general approach, MCR is currently a computationally expensive task.
This is why the aim of this article is to present an efficient approach for reconstructing three-dimensional (3D) microstructures from universal descriptors.
A brief introduction to MCR and the current state of the art is given in the following.

The microstructure, hereafter also referred to as structure, which lies at the heart of the process-structure-property linkages, poses a fundamental problem:
Due to the morphological randomness of a material, two random realizations of the same structure usually look similar, but correspond to very different locations in pixel space, i.e. the exact location and shape of structural features differ. 
For operations like quantitatively comparing microstructures or even interpolating between them, the microstructure needs to be mapped to a \textit{translation-invariant, stationary descriptor}~$\boldsymbol{D}$ that allows for these operations.
These properties of~$\boldsymbol{D}$ provide a sound foundation for representing structures in process-structure and structure-property linkages. Furthermore, it provides a space for sampling synthetic structures, thus enabling data-driven materials development workflows.

Such a descriptor $\boldsymbol{D}$ can be composed of classical descriptors like volume fractions, grain size distributions and grain aspect ratio distributions~\cite{xu_descriptor-based_2014}. 
In the general case, however, high-dimensional statistical descriptors are preferred, like the lineal path function~\cite{lu_lineal-path_1992}, cluster correlation function~\cite{jiao_superior_2009} or spatial $n$-point correlations~$\boldsymbol{S}$~\cite{jiao_modeling_2007}.
Especially the latter one is widely used~\cite{kalidindi_microstructure_2011}.
Integral descriptors like Minkowski tensors have also received considerable attention~\cite{scheunemann_design_2015}.
Furthermore, a direct description of the underlying random field can yield microstructure descriptors. Examples are weakly non-uniform Gaussian random fields~\cite{liu_translation_2016,liu_direct_2017} and an explicit mixture random field~\cite{gao_ultra-efficient_2021}.
Recently, descriptors have also been learned from data. In the various recently proposed data-driven MCR approaches, e.g.\ \cite{yang_microstructural_2018, kench_generating_2021, zhang_slice--voxel_2021}, to name only a few, the descriptors are not formulated explicitly, but arise during model training as a latent vector in the network.
For this purpose, the training data set does not need to be structure-specific. This has been demonstrated in~\cite{lubbers_inferring_2017}, where the Gram matrices of a pre-trained convolutional neural network are identified as viable microstructure descriptors.
In summary, the literature comprises a great variety of descriptors to quantitatively characterize microstructures for establishing process-structure-property linkages as well as for generating computational domains to conduct numerical simulations. Developing new descriptors is an active field of research.
A central question is therefore how to reconstruct a microstructure given a descriptor.

Microstructure reconstruction is reviewed in~\cite{bostanabad_computational_2018} and briefly summarized in the following.
The Yeong-Torquato algorithm is a well-known and universal approach that allows to reconstruct a structure given \emph{any} descriptor~\cite{yeong_reconstructing_1998}.
Unfortunately, for high requirements regarding the reconstruction accuracy and the microstructure resolution, the algorithm scales highly unfavorably.
For this reason, various accelerated algorithms have been proposed that trade the universality for a gain in speed.
For example, in~\cite{scheunemann_design_2015}, a matrix structure with arbitrarily shaped and spatially distributed inclusions is approximated through a small number of elliptical inclusions, accomplishing high performance for microstructures that fit this ansatz.
As a second example, the well-known software \emph{DREAM.3D}~\cite{groeber_dream3d_2014} achieves very high performance by both using a feature-based microstructure ansatz and fixing the prescribable descriptors to classical variants like grain size distributions.
Restricting not the microstructure itself, but only the prescribable descriptors, various data-driven reconstruction algorithms have been proposed based on Generative Adversarial Networks~(GANs)~\cite{yang_microstructural_2018, kench_generating_2021}, sometimes in combination with Variational Autoencoders~(VAEs)~\cite{zhang_slice--voxel_2021,noguchi_stochastic_2021}. These approaches achieve the highest speed in the literature, but require a training data set and can only reconstruct from a given point in the latent space\footnote{The internal representation of an image in a neural network is defined as point in the latent space. Its variables encode the associated image in an abstract and usually condensed manner. The distinct meaning of these variables cannot be directly determined as it results from the chosen architecture and the training process.}, not from a chosen descriptor.
In absence of a training data set, the transfer learning approach from~\cite{li_transfer_2018} becomes interesting, which has recently been extended to 3D~\cite{bostanabad_reconstruction_2020}. 
It makes no assumptions regarding microstructure, but the descriptor is restricted to Gram matrices\footnote{With the activation of the $p$-th channel of a convolutional neural network in layer $l$ at the spatial position $i$ denoted as $F_{ip}^l$, the Gram matrix of layer $l$ is expressed as $G_{pr}^l = \sum_i F_{ip}^l F_{ir}^l$.}.
An extension to combine the approach with spatial two-point correlations is given in~\cite{bhaduri_efficient_2021}.
Finally, extremely efficient reconstructions without training on a specific data set are also obtained by limiting the descriptor to the underlying random field~\cite{gao_ultra-efficient_2021}. In this case, the quality of the reconstructed microstructures depends on how well this descriptor captures the relevant microstructural features.
In summary, despite the variety of different approaches for MCR, a fast and flexible method for arbitrary descriptors is still missing.

The objective of this work is to present an efficient general-purpose 3D microstructure reconstruction framework to combine the generality of the Yeong-Torquato algorithm with the speed of new methods without requiring a data set.
The solution is based on \emph{differentiable MCR}~\cite{seibert_reconstructing_2021}, a two-dimensional (2D) algorithm that makes no restrictions for the microstructure and allows any descriptor as long as it is differentiable.
For real-world applications, however, the same efficiency and flexibility needs to be cast to a fully 3D formulation.
Thereby, the main challenges are not only to retain speed and scalability, but also to address potential noise, which is known to be more problematic for 3D than for 2D~\cite{bostanabad_reconstruction_2020}.
Furthermore, a descriptor-based reconstruction algorithm must meet the experimental limitation that real microstructure data in the form of microscopy images is commonly available on 2D slices only. The objective of the present work is to meet these challenges in the framework of \emph{differentiable MCR}.

The article begins by presenting the formulation in \autoref{sec:formulation}. Starting from a review of \emph{differentiable MCR} in \autoref{ssec:dmcr}, a slice-based extension to 3D is presented in \autoref{ssec:slices}. A fundamental difference to the literature lies in how the emerging noise is handled in \autoref{ssec:variation}. Results are shown and discussed in \autoref{sec:results} and a conclusion is drawn in \autoref{sec:conclusions}.

The following notation is used: Upper and lower case letters~$(i, \, I)$ indicate scalars, and bold letters~$(\boldsymbol{M}, \boldsymbol{D})$ represent arrays with scalar entries~$({M}_{i,j,k}, {D}_{k})$. The modulo operator~$\text{mod}_I (i)$ yields the remainder of the integer division of~$i$ by~$I$. For example,~$\text{mod}_4 (9) = 1$. The treatment of negative~$i$ is defined by requiring~$\text{mod}_I (i) = \text{mod}_I (i + nI) \; \forall \; n \in \mathbb{Z}$.

\section{Methods}\label{sec:formulation}

\subsection{Differentiable MCR in 2D}\label{ssec:dmcr}
Given the desired microstructure descriptor~$\boldsymbol{D}^\text{des}$, the goal of microstructure reconstruction is to find a microstructure~$\boldsymbol{M}$ such that the corresponding descriptor~$\boldsymbol{D}(\boldsymbol{M})$ is as close as possible to the desired value~$\boldsymbol{D}^\text{des}$.
Consequently, it is reasonable to formulate microstructure reconstruction as an optimization problem.
In order to render this optimization procedure efficient, \emph{differentiable MCR} has been proposed for 2D in~\cite{seibert_reconstructing_2021} and a brief introduction is given in the following. 

The central paradigm of \emph{differentiable MCR} is that microstructure reconstruction is optimization and that efficient optimization requires the gradient.
This is expressed as
\begin{equation}\label{eq:dMCR-Min_Prob}
    \underset{\boldsymbol{\tilde{M}}}{\mathrm{argmin}} \; \mathcal{L}(\boldsymbol{\tilde{M}}) \; ,
\end{equation}
where each pixel of the intermediate microstructure $\boldsymbol{\tilde{M}}$ is allowed to take any real value in the interval~$(0, \, 1)$ during the reconstruction process. Only after the reconstruction is done,~$\boldsymbol{\tilde{M}}$ is rounded to obtain an integer-valued microstructure~$\boldsymbol{M}$, where each pixel is either~$0$ or~$1$.
The loss function 
\begin{equation}\label{eq:L_func_dMCR}
	\mathcal{L}(\boldsymbol{\tilde{M}}) =  \left|\left| \boldsymbol{\tilde{D}}(\boldsymbol{\tilde{M}}) - \boldsymbol{D}^{\text{des}} \right|\right| \; 
\end{equation}
quantifies the difference between the desired descriptor~$\boldsymbol{D}^\text{des}$ and the descriptor that is associated with the current intermediate microstructure~$\boldsymbol{\tilde{M}}$. 
\begin{figure*}[t]
    \centering
    \includegraphics[width = \textwidth]{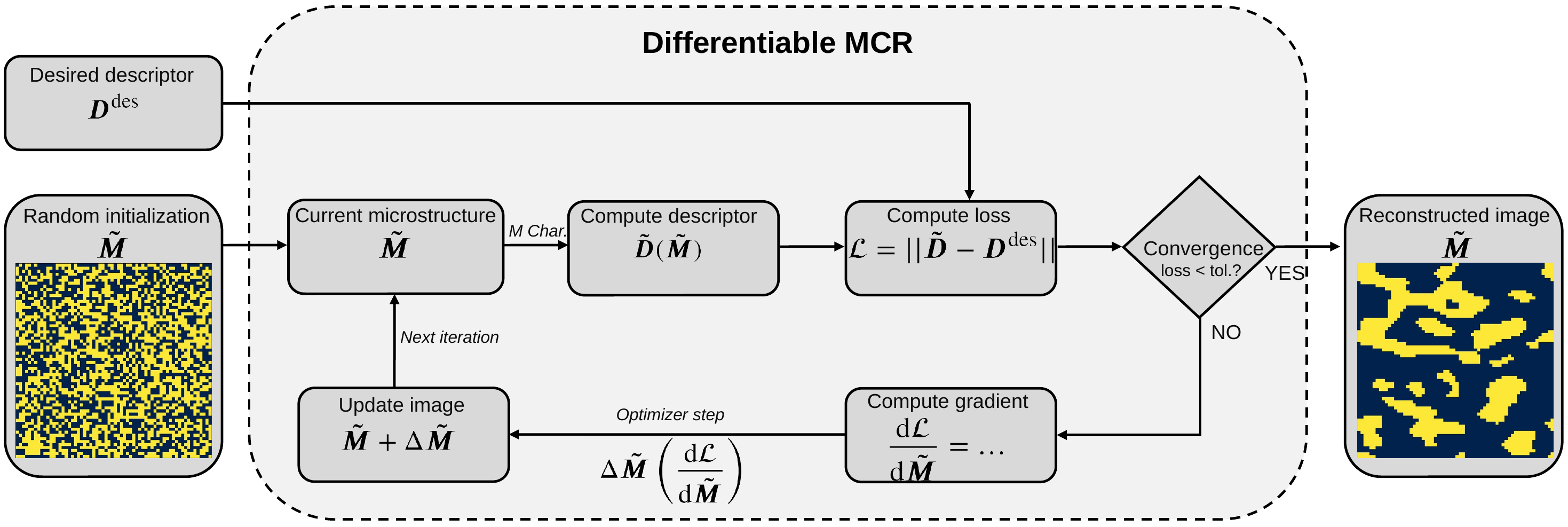}
    \caption{Flow chart illustrating the proposed differentiable microstructure characterization and reconstruction (MCR), reproduced from~\cite{seibert_reconstructing_2021} with kind permission from the publisher.\label{fig:flowchart}}
\end{figure*}

The optimization problem~\autoref{eq:dMCR-Min_Prob} is solved in an iterative manner as illustrated in~\autoref{fig:flowchart}.
In each iteration, the microstructure~$\boldsymbol{\tilde{M}}$ is altered in order to decrease the loss~$\mathcal{L}$.
In order to solve this optimization problem quickly, the change in microstructure~$\Delta \boldsymbol{\tilde{M}}$ is computed\footnote{For this, any gradient-based optimizer can be used, for example the ADAM-optimizer or a quasi-Newton method like L-BFGS-B.} using the gradient~$\text{d}\mathcal{L}/\text{d}\boldsymbol{\tilde{M}}$.
This requires~$\mathcal{L}$ to be differentiable with respect to~$\boldsymbol{\tilde{M}}$. 
With \autoref{eq:L_func_dMCR}, this is fulfilled if the same is true for the descriptor. 
Therefore, we introduce the differentiable descriptor~$\boldsymbol{\tilde{D}}(\boldsymbol{\tilde{M}})$, which is defined and differentiable for each possible intermediate microstructure~$\boldsymbol{\tilde{M}}$.
Furthermore, the special case of an integer-valued microstructure should be reproduced:
$$\boldsymbol{\tilde{D}}(\boldsymbol{M}) = \boldsymbol{D}(\boldsymbol{M}) \;.$$
Examples for such differentiable descriptors are the Gram matrices~$\boldsymbol{G}$ from~\cite{lubbers_inferring_2017,li_transfer_2018,bostanabad_computational_2018} and the differentiable generalization of spatial $n$-point correlations~$\boldsymbol{S}$ that has been proposed in~\cite{seibert_reconstructing_2021}. Therefore, some recent and potent reconstruction algorithms can be identified as special cases of \emph{differentiable MCR}~\cite{li_transfer_2018,bostanabad_reconstruction_2020,bhaduri_efficient_2021}. 

\begin{figure}[t]
	\centering
    	\includegraphics[width=0.8\linewidth]{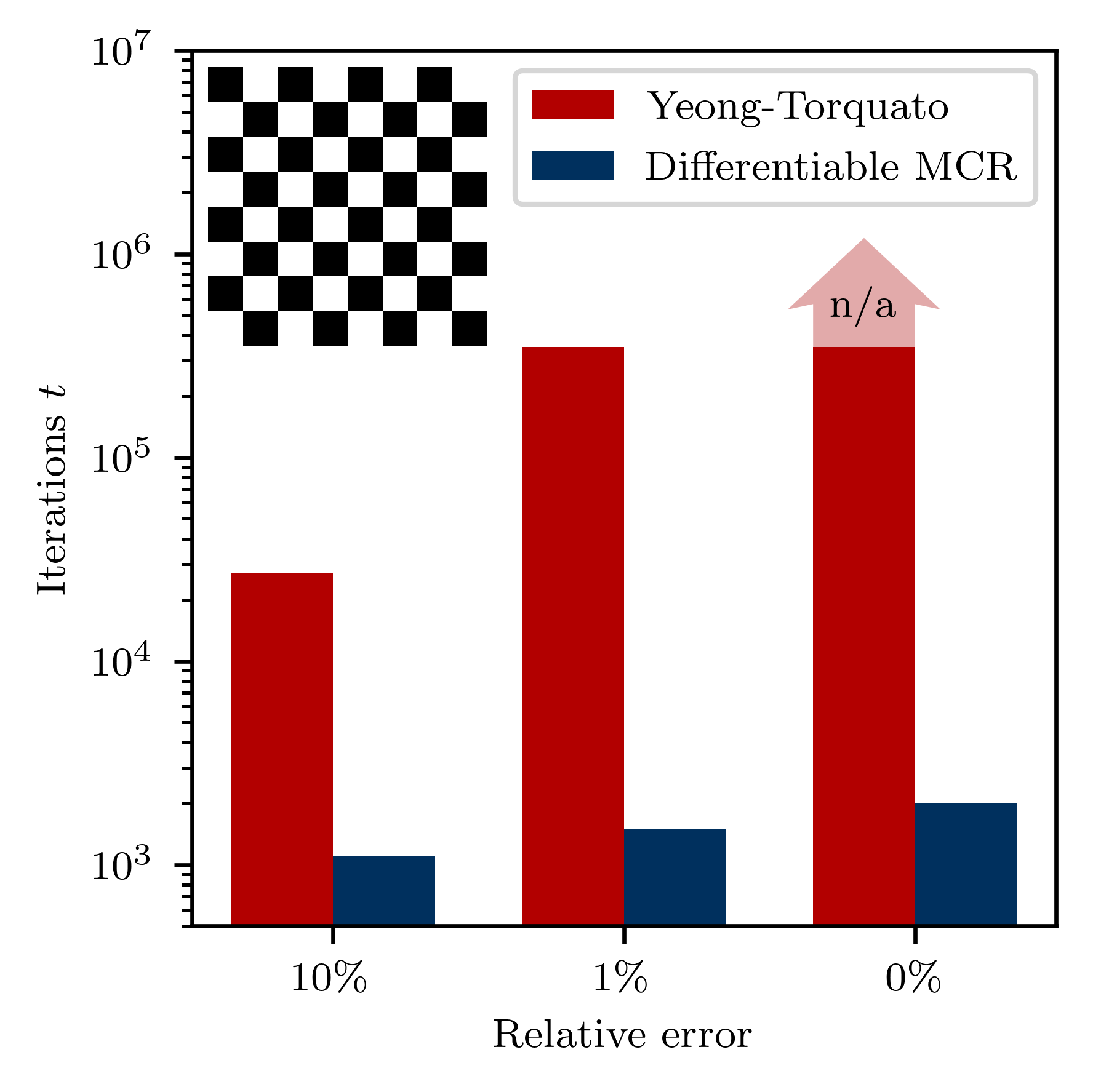}
	\caption[Barplot]{Iterations needed for reconstructing a checkerboard pattern in~\cite{seibert_reconstructing_2021}. Differentiable MCR requires less iterations than the Yeong-Torquato algorithm and scales better as the error decreases. After 2000 iterations, an error of~$0\%$ is reached, which is not achievable with the Yeong-Torquato algorithm in a feasible time.\label{fig:barplot}}
\end{figure}

Since the proposed microstructure reconstruction is formulated as a \emph{gradient-based} optimization procedure, \emph{differentiable MCR} significantly reduces the computational costs in comparison with the standard Yeong-Torquato algorithm, as shown in \autoref{fig:barplot} and in~\cite{seibert_reconstructing_2021}. 
Furthermore, it scales more favorably with high resolution and accuracy requirements.
With \emph{differentiable MCR} being formulated in~\cite{seibert_reconstructing_2021} only in 2D, this superior efficiency and scaling motivate an extension to 3D, which is presented in the next section.

\subsection{Slicing-based generalization to 3D}\label{ssec:slices}
Inspired by the work of Bostanabad~\cite{bostanabad_reconstruction_2020}, the proposed \emph{differentiable MCR} algorithm is formulated in 3D as illustrated in Figure~\ref{fig:slicing}: 
\begin{figure*}[t]
	\centering
	\subfloat[3D Microstructure]{\includegraphics[height=0.24\textwidth]{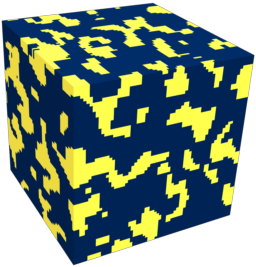}}
	\hfill
	\subfloat[Slicing in x-direction]{\includegraphics[height=0.24\textwidth]{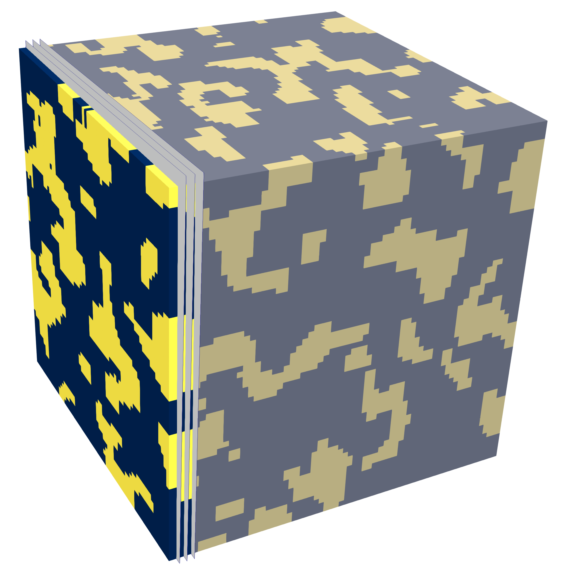}}
	\hfill
	\subfloat[Slicing in y-direction]{\includegraphics[height=0.24\textwidth]{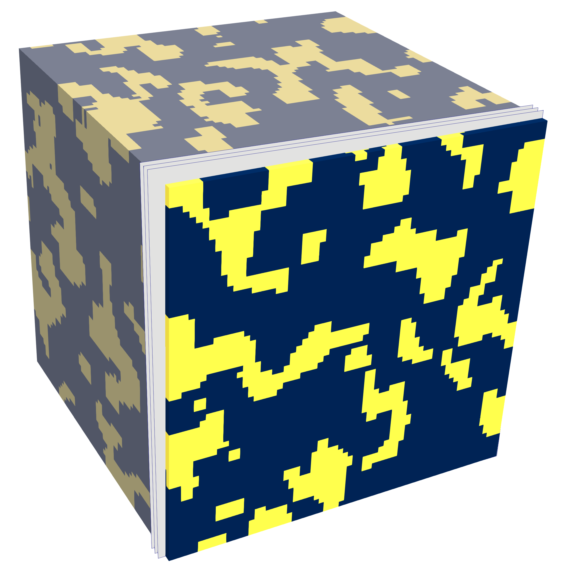}}
	\hfill
	\subfloat[Slicing in z-direction]{\includegraphics[height=0.24\textwidth]{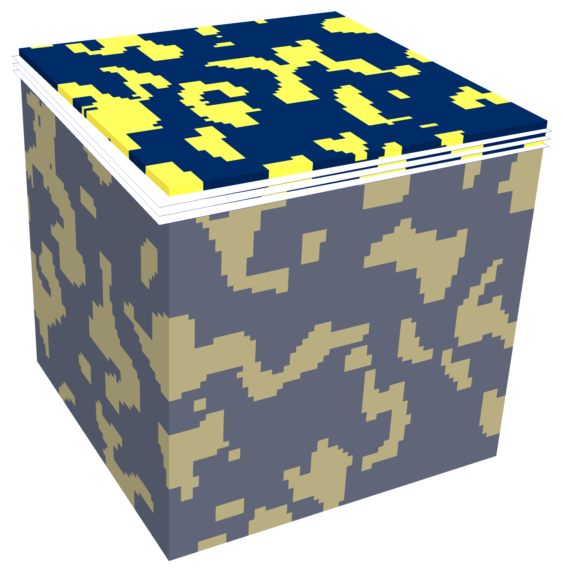}}
	\hfill
	\caption[Sliced microstructure]{An exemplary microstructure of~$64 \times 64 \times 64$ voxels is sliced in each spatial dimension to obtain a total of~$3 \times 64$ slices, each with~$64 \times 64$ pixels.\label{fig:slicing}}
\end{figure*}
Instead of computing descriptors directly in~3D, the microstructure is sliced in all three spatial directions\footnote{In principle, any directions can be chosen. As a simple and intuitive example, this work uses orthogonal sections that align with the specimen shape.} to obtain stacks of 2D microstructure sections. The descriptor can then be computed for each slice separately. This is especially expedient as microstructure samples are commonly available as 2D microscopy images only. Therefore, on each slice, the descriptor is required to resemble the desired~2D descriptor as closely as possible. This allows to consider anisotropy, as different descriptors can be prescribed along different directions. This leads to minimizing the following loss function
\begin{align}\label{eq:slicinganiso}
    \nonumber \mathcal{L} = & \sum_{i}^{n_x^\text{slices}} \left|\left| \boldsymbol{\tilde{D}}_i^\text{x}(\boldsymbol{\tilde{M}}) -  \boldsymbol{D}^{\text{des, x}} \right|\right| + \\
    \nonumber + & \sum_{j}^{n_y^\text{slices}} \left|\left| \boldsymbol{\tilde{D}}_j^\text{y}(\boldsymbol{\tilde{M}}) -  \boldsymbol{D}^{\text{des, y}} \right|\right| + \\
    + & \sum_{k}^{n_z^\text{slices}} \left|\left| \boldsymbol{\tilde{D}}_k^\text{z}(\boldsymbol{\tilde{M}}) -  \boldsymbol{D}^{\text{des, z}} \right|\right| \; ,
\end{align}
where~$\boldsymbol{\tilde{D}}_i^\text{x}(\boldsymbol{\tilde{M}})$ returns the 2D descriptor of the $i$-th slice of~$\boldsymbol{\tilde{M}}$ in x-direction,  $\boldsymbol{\tilde{D}}_j^\text{y}(\boldsymbol{\tilde{M}})$ yields the 2D descriptor of the $j$-th slice of~$\boldsymbol{\tilde{M}}$ in y-direction and $\boldsymbol{\tilde{D}}_k^\text{z}(\boldsymbol{\tilde{M}})$ represents the 2D descriptor of the $k$-th slice of~$\boldsymbol{\tilde{M}}$ in z-direction. The $\boldsymbol{D}^{\text{des, x}}, \boldsymbol{D}^{\text{des, y}}$ and $\boldsymbol{D}^{\text{des, z}}$ are the desired~2D descriptor in the $x,y$ and $z$ directions, respectively\footnote{In principle, a spatially varying desired descriptor can be considered, e.g. for achieving a functionally graded microstructures.}.
In the general anisotropic case,
$$\boldsymbol{D}^{\text{des, x}} \neq \boldsymbol{D}^{\text{des, y}} \neq \boldsymbol{D}^{\text{des, z}} \;$$
holds.
If, however, isotropic materials are considered, i.e.
$$\boldsymbol{D}^{\text{des, x}} = \boldsymbol{D}^{\text{des, y}} = \boldsymbol{D}^{\text{des, z}} =: \boldsymbol{D}^{\text{des}}\; ,$$
then \autoref{eq:slicinganiso} simplifies to
\begin{equation}\label{eq:slicingiso}
    \mathcal{L} = \sum_s^{n^\text{slices}} \left|\left| \boldsymbol{\tilde{D}}_s(\boldsymbol{\tilde{M}}) -  \boldsymbol{{D}}^{\text{des}} \right|\right| \; .
\end{equation}
This formulation reproduces the special case of \autoref{eq:L_func_dMCR} if only one slice is present.

Even though such an extension is conceptually simple and universally applicable, minimizing \autoref{eq:slicinganiso} or \autoref{eq:slicingiso} leads to noisy results, as visible in \autoref{fig:3dcomparison-a}.
The noise, which is not present in 2D-reconstructions~\cite{seibert_reconstructing_2021}, is presumably caused by the slice-based approach, as the data only allows for an incomplete descriptor specification.
In order to solve this problem, a common denoising strategy is introduced and ameliorated in the next section.

\subsection{Noise handling using total variation}\label{ssec:variation}
Noise is a major hurdle when reconstructing 3D microstructures from 2D slices.
This is solved in the literature using the total variation, which is presented in \autoref{sssec:tvd}.
The common way of using the total variation during the reconstruction process is described in \autoref{ssec:tv_common}, however, a more rigorous and accurate approach is developed in \autoref{ssec:tv_we}.

\subsubsection{Original technique: Total variation denoising}\label{sssec:tvd}
Total variation denoising (TVD) is an image denoising technique~\cite{rudin_nonlinear_1992} that can be readily applied to microstructure reconstruction as a post-processing step without any significant adaption of the formulation. It is illustrated in \autoref{fig:tvd}:
\begin{figure}[t]
	\centering
	\begin{subfigure}[b]{0.32\linewidth}
	    \centering
    	\includegraphics[width=\linewidth]{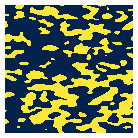}
    	\caption{Original image\label{fig:tvd-a}}
	\end{subfigure}
	\hfill
	\begin{subfigure}[b]{0.32\linewidth}
	    \centering
    	\includegraphics[width=\linewidth]{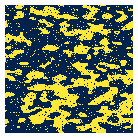}
    	\caption{Noisy version\label{fig:tvd-b}}
	\end{subfigure}
	\hfill
	\begin{subfigure}[b]{0.32\linewidth}
	    \centering
    	\includegraphics[width=\linewidth]{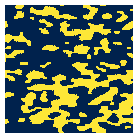}
    	\caption{Denoised image\label{fig:tvd-c}}
	\end{subfigure}

	\caption[Total variation denoising]{Illustration of total variation denoising~\cite{rudin_nonlinear_1992}: Given a noisy image (b) a noise-free approximation (c) to the original image (a) is found by prescribing proximity to the noisy image while at the same time minimizing the variation. The original microstructure is generated synthetically using~\emph{pyMKS}~\cite{brough_materials_2017}.\label{fig:tvd}}
\end{figure}
Starting with a noisy version of an image (\autoref{fig:tvd-b}), the goal of denoising is to produce a denoised image (\autoref{fig:tvd-c}) which is as close as possible to the original noise-free image (\autoref{fig:tvd-a}). For this purpose, the level of noise in an image is measured via the variation. The local variation at each pixel is its absolute difference in value compared to the neighboring\footnote{Only two of the four neighbors need to be considered because of the spatial symmetry of the operation.} pixels. 
This variation is accumulated over all pixels to obtain the total variation~$\mathcal{V}$. 
In~2D, it can be computed as
$$
    \mathcal{V}^\text{2D} = \sum_{i, j}^{I,J} \left[ \left| M_{i, j} - M_{\text{mod}_I(i-1), j} \right| + \left| M_{i, j} - M_{i, \text{mod}_J(j-1)} \right| \right] \; ,
$$
where the modulo operator reflects the periodicity of the microstructure. In 3D, the equation extends to
\begin{align}\label{eq:msvariation}
    \nonumber \mathcal{V}^\text{3D} = \sum_{i, j, k}^{I,J,K} \bigg[ & \left| M_{i, j, k} - M_{\text{mod}_I(i-1), j, k} \right| +  \\ 
    \nonumber + & \left| M_{i, j, k} - M_{i, \text{mod}_J(j-1), k} \right| + \\
    + & \left| M_{i, j, k} - M_{i, j, \text{mod}_K(k-1)} \right| \bigg] \; ,
\end{align}
where $I$, $J$ and $K$ denote the number of pixels in $x$, $y$ and $z$ direction, respectively.

With this definition, it becomes clear that the original image has in \autoref{fig:tvd} has a lower variation than a noisy version of it. 
Based on this observation, TVD is formulated as the convex optimization problem 
\begin{equation}\label{eq:tvd}
    \boldsymbol{M}^{\text{TVD}} = \underset{\boldsymbol{M}}{\textrm{argmin}} \left[ \left| \left| \boldsymbol{M} - \boldsymbol{M}^{\text{noise}} \right|  \right| + \lambda \cdot \mathcal{V}(\boldsymbol{M}) \right] \; .
\end{equation}
The denoised image~$\boldsymbol{M}^{\text{TVD}}$ should be as close as possible to the available noisy image, but also have a low variation. As these two goals contradict each other, a parameter~$\lambda$ is introduced as a trade-off. As~$\lambda \to 0$, it is clear that~\mbox{$\boldsymbol{M}^{\text{TVD}} \to \boldsymbol{M}^{\text{noise}}$}, whereas for~$\lambda \to \infty$, $\boldsymbol{M}^{\text{TVD}}$ becomes a uniform image with~$\mathcal{V} = 0$. An intermediate~$\lambda$ is required for obtaining a result as shown in \autoref{fig:tvd}. Despite the necessity to find the weight~$\lambda$ for the variation penalty term, TVD is especially well suited for denoising microstructures: Unlike other common denoising strategies like filter-based smoothing, TVD retains sharp borders.

The approach in \autoref{eq:tvd} contains the microstructure directly and not in the form of a descriptor. Furthermore, it requires a noisy MS to start with. It is therefore really decoupled from the reconstruction and merely a post-processing step. In practice, however, it is not common to use TVD as a post-processing step after microstructure reconstruction, but rather to add the variation as a regularization term during the reconstruction itself as described next.

\subsubsection{Established approach: Minimizing the total variation during reconstruction}\label{ssec:tv_common}
In order to suppress noise during the slice-based reconstruction of a 3D microstructure, {Bostanabad}~\cite{bostanabad_reconstruction_2020} proposed to add the total variation~$\mathcal{V}$ to the reconstruction loss function
\begin{equation}\label{eq:3dbostanabad}
    \mathcal{L} = \sum_s^{n^\text{slices}} \left|\left| \boldsymbol{\tilde{D}}_s(\boldsymbol{\tilde{M}}) -  \boldsymbol{{D}}^{\text{des}} \right|\right| + \lambda \mathcal{V}^\text{3D}(\boldsymbol{\tilde{M}}) \; .
\end{equation}
A similar formulation has been used in~\cite{bhaduri_efficient_2021}.
This avoids the necessity of a post-processing step like \autoref{eq:tvd} that introduces errors in the descriptor.
In~\cite{bostanabad_reconstruction_2020}, only Gram matrices are used as a descriptor, but in fact, \emph{any} descriptor can be used. Thus, the generality of \emph{differentiable MCR} is retained.

While significantly reducing the noise in reconstructed microstructures, the total variation penalty term of \autoref{eq:3dbostanabad} introduces two additional problems:
\begin{enumerate}
	\item The weight $\lambda$ needs to be tuned by hand, introducing manual and computational effort.
	\item No difference is made between noise and structure. For example, to the optimizer, spurious noise and real pores both are sources of variation and need to be kept at a minimum. Hence, the denoising term stands in direct conflict with the descriptor error term. If~$\lambda$ is large enough to suppress noise, it might also be so large that relevant microstructural features are missing and there might exist a noise-free solution which is better in terms of descriptor error, but is not reached by the optimizer because of its higher variation.
\end{enumerate}
These two problems are solved through the proposed approach presented in the next \autoref{ssec:tv_we}.

\subsubsection{Proposed approach: Total variation as a descriptor}\label{ssec:tv_we}
In the present contribution, it is proposed to reinterpret the variation as a microstructure descriptor and to match its desired value instead of minimizing it.

\begin{figure*}
	\centering
	\import{Figures/}{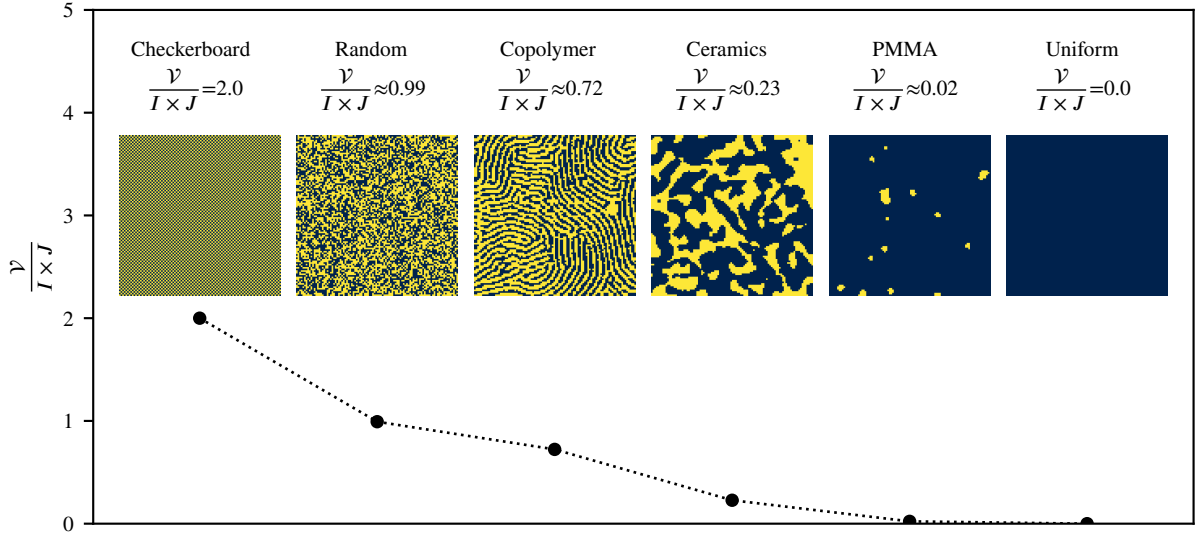}
	\caption[Variation as a descriptor]{Illustration of the interpretability of the normalized total variation~$\mathcal{V} / \left(I \times J\right)$ as a microstructure descriptor: Different structures have different variations, according to the amount of phase boundary. Some microstructures are taken from~\cite{li_transfer_2018}, re-used with kind support from the authors and permission from the publisher through the Creative Commons license~\cite{creative_commons}.\label{fig:variationasadescriptor}}
\end{figure*}

As a motivation, a noise-free and two-phase microstructure~$\boldsymbol{{M}}$ is considered. In this case, the definition of the variation is easy to interpret: In each spatial direction, the corresponding term in \autoref{eq:msvariation} is~1 if a phase boundary is present and~0 otherwise. It is therefore an indicator function for the phase boundary. The sum over all pixels approximates an integration. In a voxel mesh, the total variation~$\mathcal{V}$ is thus the discretized integral over the phase boundaries. If~$\mathcal{V}$ is properly normalized with the volume, it can serve as a microstructure descriptor quantifying the amount of phase boundary per unit volume. Such a descriptor is large for lamellae and small for equiaxed features, as shown in \autoref{fig:variationasadescriptor}. Furthermore,~$\mathcal{V}$ is defined for real-valued microstructures~$\boldsymbol{\tilde{M}}$ and can be differentiated with respect to~$\boldsymbol{\tilde{M}}$. The normalized variation is therefore a valid differentiable microstructure descriptor.

Following the paradigm of \emph{differentiable MCR},~$\mathcal{V}$ should not be minimized, but rather optimized to match the desired value. \autoref{eq:3dbostanabad} thus becomes
$$
    \mathcal{L} = \sum_s^{n^\text{slices}} \left|\left| \boldsymbol{\tilde{D}}_s(\boldsymbol{\tilde{M}}) -  \boldsymbol{D}^{\text{des}} \right|\right| + \lambda \left(\mathcal{V}_s^\text{2D}(\boldsymbol{\tilde{M}}) - \mathcal{V}^{\text{2D, des}}\right)^2 \; .
$$
Alternatively, one can think of concatenating a generic descriptor~$\boldsymbol{\tilde{D}}$ and the variation~$\mathcal{V}^\text{2D}$ as
\begin{equation}\label{eq:concatenation}
	\boldsymbol{\tilde{D}}^{\mathcal{V}} = \left( \tilde{D}_1, \tilde{D}_2, ..., \tilde{D}_K, \lambda \mathcal{V}^\text{2D} \right)
\end{equation}
and formulating the loss function in a more unified way as
\begin{equation}\label{eq:3dours}
    \mathcal{L} = \sum_s^{n^\text{slices}} \left|\left| \boldsymbol{\tilde{D}}_s^{\mathcal{V}}(\boldsymbol{\tilde{M}}) -  \boldsymbol{D}^{\mathcal{V}, \text{des}} \right|\right| \; .
\end{equation}
From this perspective, the notion of a regularization term can be replaced with the perspective that the variation is a physically motivated microstructure descriptor that is especially suitable for obtaining noise-free~3D reconstructions from~2D slices.
Finally, it is worth pointing out that the total variation is identified as a grain boundary integral \emph{only} {for a voxelized geometry}, where the integral yields a Manhattan metric. 
This is illustrated in greater detail in \autoref{sec:piis4}. 

With this reformulated loss function, all other aspects of \emph{differentiable MCR} remain unchanged: As shown in \autoref{fig:flowchart},~$\mathcal{L}$ is differentiated with respect to~$\boldsymbol{\tilde{M}}$ and minimized through a suitable optimizer.

\section{Results and discussion}\label{sec:results}
Various numerical experiments are performed in order to show the potential of the proposed approach.
First, in \autoref{ssec:results_variation}, the influence of the total variation~$\mathcal{V}$ is investigated and the common approach of minimizing~$\mathcal{V}$ is compared to the proposed approach of matching its desired value.
Secondly, the flexibility of the proposed approach regarding the choice of descriptor is demonstrated in \autoref{ssec:results_descriptorchoice}. This allows for choosing the best descriptor for each situation and to perform explicit interpolations in a chosen descriptor space.
Thirdly, in \autoref{ssec:results_perfo}, the performance and scalability of the method is investigated.
Finally, in \autoref{ssec:results_further-results}, a variety of different microstructures is reconstructed to demonstrate the range of applicability of the method.
Details on the implementation can be found in~\autoref{sec:methods}.

\subsection{Influence of variation}\label{ssec:results_variation}
In this numerical experiment, the influence of the total variation~$\mathcal{V}$ is investigated by demonstrating its relevance and comparing different formulations.
The focus thereby lies on comparing the common approach of minimizing~$\mathcal{V}$ as in \autoref{ssec:tv_common} to the proposed approach of matching its desired value as in \autoref{ssec:tv_we}.
In order to achieve comparability to Bostanabad's approach~\cite{bostanabad_reconstruction_2020}, {Gram} matrices~$\boldsymbol{G}$ are used as a descriptor in this section.

\begin{figure*}
	\begin{subfigure}[b]{0.20\linewidth}
	    \centering
    	\includegraphics[width=\linewidth]{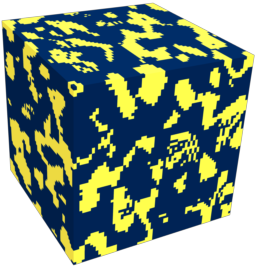}
    	\caption{Minimize~$\mathcal{V}$; $\lambda=10^{-6}$\label{fig:3dcomparison-a}}
	\end{subfigure}
	\hfill
	\begin{subfigure}[b]{0.20\linewidth}
	    \centering
    	\includegraphics[width=\linewidth]{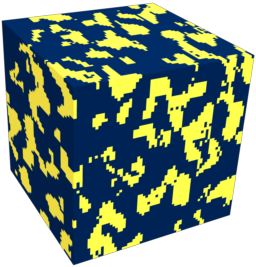}
    	\caption{Minimize~$\mathcal{V}$; $\lambda=10^{-4}$\label{fig:3dcomparison-b}}
	\end{subfigure}
	\hfill
	\begin{subfigure}[b]{0.20\linewidth}
	    \centering
    	\includegraphics[width=\linewidth]{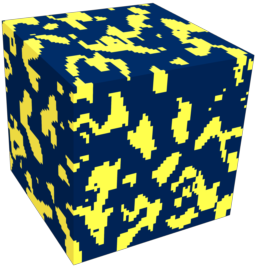}
    	\caption{Minimize~$\mathcal{V}$; $\lambda=10^{-3}$\label{fig:3dcomparison-c}}
	\end{subfigure}
	\hfill
	\begin{subfigure}[b]{0.20\linewidth}
	    \centering
    	\includegraphics[width=\linewidth]{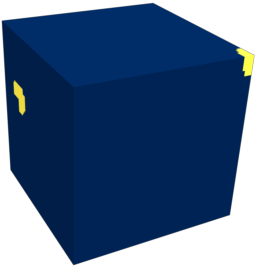}
    	\caption{Minimize~$\mathcal{V}$; $\lambda=10^{-2}$\label{fig:3dcomparison-d}}
	\end{subfigure}
	\vfill
	\begin{subfigure}[b]{0.20\linewidth}
	    \centering
    	\includegraphics[width=\linewidth]{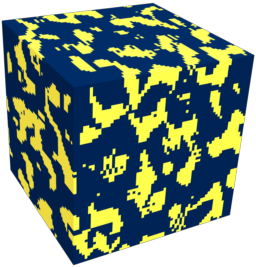}
    	\caption{Match~$\mathcal{V}$; $\lambda=10^{-2}$\label{fig:3dcomparison-e}}
	\end{subfigure}
	\hfill
	\begin{subfigure}[b]{0.20\linewidth}
	    \centering
    	\includegraphics[width=\linewidth]{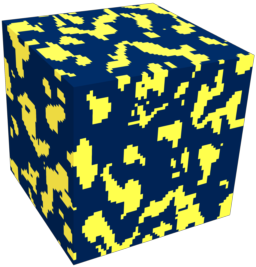}
    	\caption{Match~$\mathcal{V}$; $\lambda=10^{0}$\label{fig:3dcomparison-f}}
	\end{subfigure}
	\hfill
	\begin{subfigure}[b]{0.20\linewidth}
	    \centering
    	\includegraphics[width=\linewidth]{Figures/3d_comparison_varmat_vgg_1n.png}
    	\caption{Match~$\mathcal{V}$; $\lambda=10^{1}$\label{fig:3dcomparison-g}}
	\end{subfigure}
	\hfill
	\begin{subfigure}[b]{0.20\linewidth}
	    \centering
    	\includegraphics[width=\linewidth]{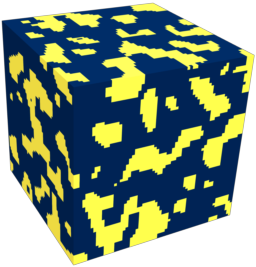}
    	\caption{Match~$\mathcal{V}$; $\lambda=10^{2}$\label{fig:3dcomparison-h}}
	\end{subfigure}
	\caption[3D reconstruction results]{3D microstructure reconstruction results based on {Gram} matrix descriptors, with the desired value derived from a 2D slice created in \emph{pyMKS}~\cite{brough_materials_2017} and shown in \autoref{fig:flowchart}. Upper row: Existing approach~\cite{bostanabad_reconstruction_2020} according to \autoref{eq:3dbostanabad}, where~$\mathcal{V}$ is minimized. Lower row: Proposed approach according to \autoref{eq:3dours}, where~$\mathcal{V}$ is matched to~$\mathcal{V}^\text{des}$.\label{fig:3dcomparison}}
\end{figure*}

\begin{figure}[t]
    \centering
    \import{Figures/}{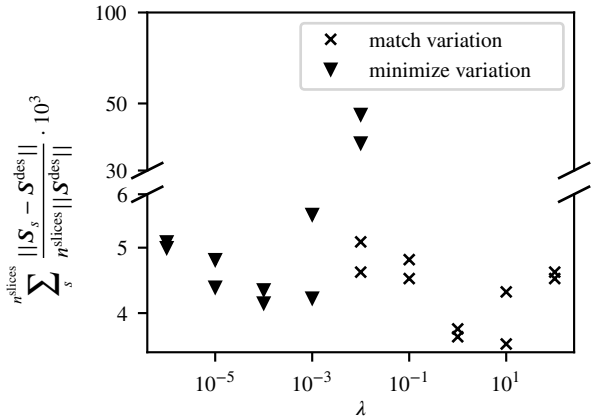}
    \caption[Error of 3D reconstruction]{Relative error of spatial two-point correlation per slice for different 3D reconstruction techniques and different variation weights~$\lambda$: For well-chosen~$\lambda$, in terms of the plotted error, the presented approach of matching the variation~$\mathcal{V}$ is superior to existing approach of minimizing~$\mathcal{V}$. Note the broken axis.\label{fig:3dnumcomparison}}
\end{figure}

A visual comparison between the approach from~\cite{bostanabad_reconstruction_2020} in \autoref{eq:3dbostanabad} and the proposed alternative in \autoref{eq:3dours} is given in \autoref{fig:3dcomparison}.
It can be seen that a higher value of~$\lambda$ yields less noise in the microstructure. However, with \autoref{eq:3dbostanabad}, key microstructural features disappear if~$\lambda$ is chosen too high.
Clearly, the proposed approach of matching the total variation~$\mathcal{V}$ instead of minimizing it is more robust\footnote{Nevertheless, if~$\lambda>10$, the procedure was found to converge much slower than with a lower~$\lambda$. In this context, it might be promising to reformulate \autoref{eq:3dours} as a constrained optimization problem via the {Lagrange} multiplier method.} with respect to calibrating the variation weight~$\lambda$. Furthermore, if~$\lambda$ is calibrated correctly, the presented approach not only produces less noise, but also reaches lower errors. This is backed up numerically in \autoref{fig:3dnumcomparison}, where the relative error per slice of the spatial two-point correlation~$\boldsymbol{S}_2$ is compared. 

\subsection{Flexible choice of descriptor}\label{ssec:results_descriptorchoice}
Rather than fixing the descriptor to a specific choice like spatial correlations, the present formulation allows to use any differentiable descriptor.
The utility of this is illustrated in Figure~\ref{fig:flexible_descriptor}:
\begin{figure*}[t]
	\centering
	\subfloat[Copolymer, $\boldsymbol{D} = \boldsymbol{S}$]{\includegraphics[width=0.2\textwidth]{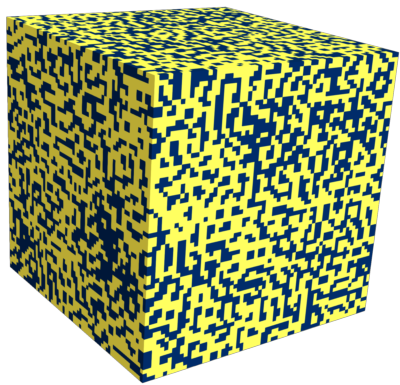}}
	\hfill
	\subfloat[Copolymer, $\boldsymbol{D} = \boldsymbol{G}$]{\includegraphics[width=0.2\textwidth]{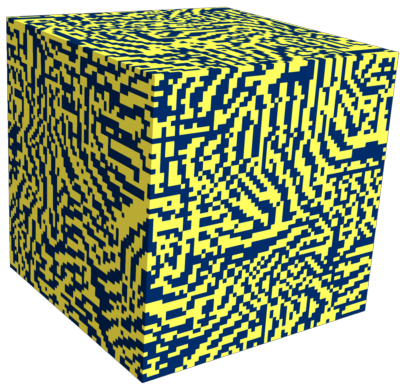}}
	\hfill
	\subfloat[Checkerboard, $\boldsymbol{D} = \boldsymbol{S}$]{\includegraphics[width=0.2\textwidth]{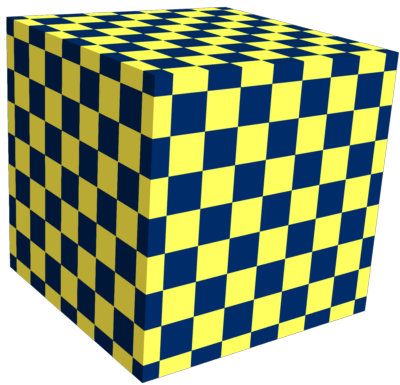}}
	\hfill
	\subfloat[Checkerboard, $\boldsymbol{D} = \boldsymbol{G}$]{\includegraphics[width=0.2\textwidth]{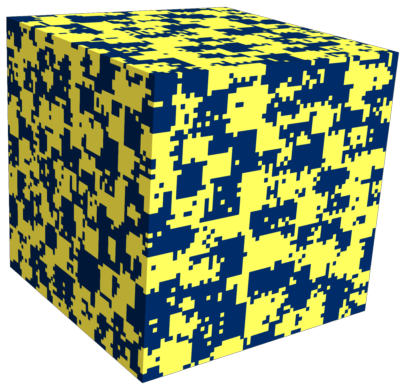}}
	\caption{Advantage of flexible descriptor choice: While Gram matrices~$\boldsymbol{G}$ outperform spatial correlations~$\boldsymbol{S}$ in reconstructing the copolymer~(a,b) from~\cite{li_transfer_2018}, the opposite is the case for a simple checkerboard structure~(c,d).\label{fig:flexible_descriptor}}
\end{figure*}
Spatial correlations fail to provide a good reconstruction of the copolymer, whereas Gram matrices perform much better.
For a simple checkerboard pattern, however, the opposite is the case.
While this is merely an academic example, the authors believe that most likely, there exists no single perfect descriptor for all possible microstructures, and that this is true for both classical statistical functions and modern data-driven descriptors.
Thus, the flexibility of the present formulation makes the method applicable to a wide range of situations.

\begin{figure*}[t]
	\centering
	\subfloat[]{\import{Figures/}{d_1.pgf}}
	\hfill
	\subfloat[]{\import{Figures/}{d_2.pgf}}
	\hfill
	\subfloat[]{\import{Figures/}{d_3.pgf}}
	\hfill
	\subfloat[]{\import{Figures/}{d_4.pgf}}
	\hfill
	\subfloat[]{\import{Figures/}{d_5.pgf}}
	\vfill
	\subfloat[]{\includegraphics[width=0.19\textwidth]{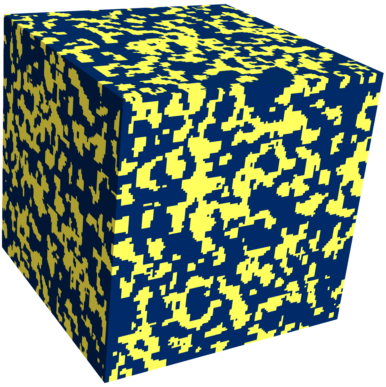}}
	\hfill
	\subfloat[]{\includegraphics[width=0.19\textwidth]{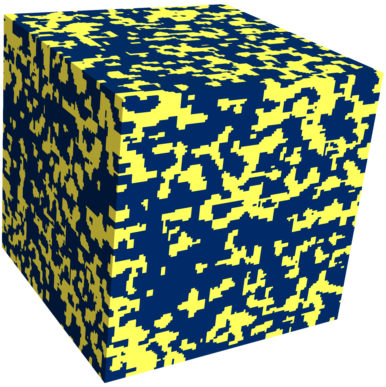}}
	\hfill
	\subfloat[]{\includegraphics[width=0.19\textwidth]{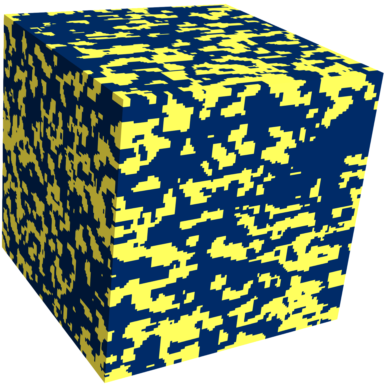}}
	\hfill
	\subfloat[]{\includegraphics[width=0.19\textwidth]{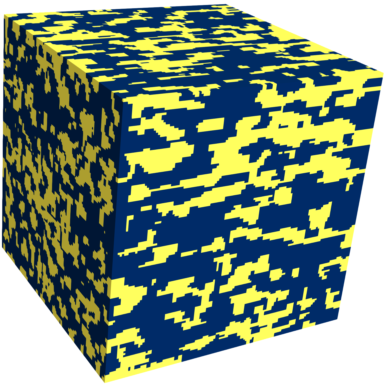}}
	\hfill
	\subfloat[]{\includegraphics[width=0.19\textwidth]{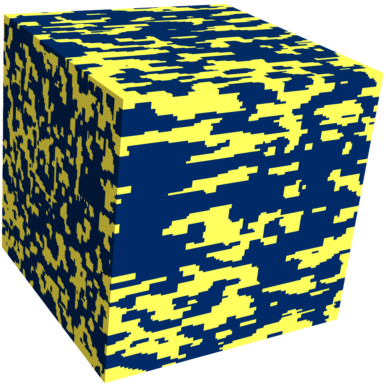}}
	\caption{Linear interpolation in descriptor space leads to smooth transitioning between microstructures~\cite{brough_materials_2017}. The descriptor space of two- and three-point correlations~$\boldsymbol{S}$ is used and visualized schematically in~(a-e) via the the two-point correlation~$\boldsymbol{S}_2$ over the distance in~$x$~($r_1$) and~$y$~($r_2$). The reconstructed structures~(f-j) are shown under the corresponding descriptor.\label{fig:interpolation}}
\end{figure*}

The explicit incorporation of the descriptor allows for interpolating between microstructures as shown in Figure~\ref{fig:interpolation}: Linearly interpolating in the descriptor space and reconstructing from these values leads to a gradual transition from isotropic to elongated structures. This allows the generation of new, unseen but feasible microstructures. In the future, these can help establishing microstructure-property linkages and optimizing microstructures.

\subsection{Performance and scalability}\label{ssec:results_perfo}
The time it takes to reconstruct a 3D microstructure mainly depends on the chosen descriptor and resolution. 
Some examples are shown in~\autoref{tab:runtimes}, although it should be added that the present code is far from fully optimized.
Because some microstructures require less iterations than others, timings are only given in an approximate interval. 
The large structure is reconstructed only once, hence, no interval can be given.
Generally, spatial correlations are observed to be more computationally efficient than Gram matrices.
Nevertheless, it can be worth combining both descriptors in order to widen the range of applicability, as discussed in \autoref{ssec:results_descriptorchoice}.
The time reported for~$\boldsymbol{G}$ without~$\boldsymbol{S}$ is slightly longer than the value reported in~\cite{bostanabad_reconstruction_2020}, although in this special case, the formulation is very similar and the hardware is identical. 
This might be attributed to the double precision used in this work, see \autoref{sec:methods}. Furthermore, in this work, periodic microstructures are created by introducing periodicity to all descriptors. 
This is computationally more expensive than non-periodic descriptors like in~\cite{bostanabad_reconstruction_2020}.
Overall, the performance of the proposed method exceeds that of the Yeong-Torquato algorithm, but does not reach the speed of random field- or GAN-based reconstruction techniques.
This is acceptable, as the reconstruction time lies in a similar order of magnitude as that of an inelastic Finite Element simulation of the microstructure.

\begin{figure*}[t]
    \centering
    \includegraphics[width=0.7\textwidth]{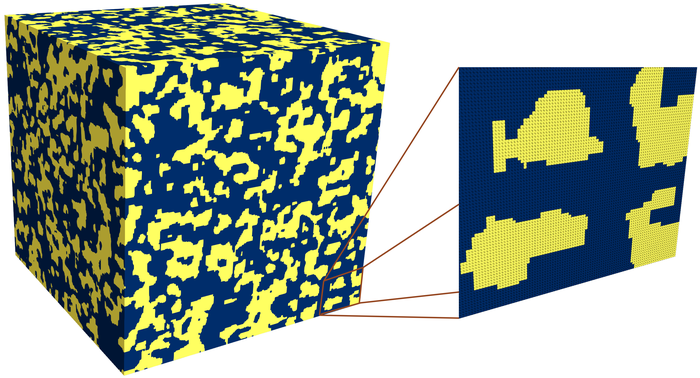}
    \caption{Reconstructed synthetic microstructure~\cite{brough_materials_2017} with $512^3 > 130{,}000{,}000$ voxels.\label{fig:large}}
\end{figure*}

To demonstrate the scalability of the presented method, \autoref{fig:large} shows a large reconstructed microstructure. 
With the multigrid scheme described in~\cite{seibert_reconstructing_2021} and seven multigrid layers, no further adaption is required.
The only bottleneck preventing a larger structure is the required memory, which the utilized code is not optimized for.
The resolution of \mbox{$512^3 > 130{,}000{,}000$}~voxels exceeds the current capabilities of GAN-based algorithms by far.
\begin{center}
    \begin{table}[t]%
        \caption{Exemplary reconstruction times for different descriptors and resolutions.\label{tab:runtimes}}
        \centering
        \begin{tabular*}{\linewidth}{@{\extracolsep\fill}lcccc@{\extracolsep\fill}}
            \toprule
            Descriptor & Size & Time & Iterations & Figure \\
            \midrule
            $\boldsymbol{S}$, $\mathcal{V}$ & $64^3$ & $20\dots 25$ m & 1000 & \ref{fig:flexible_descriptor} \\
            $\boldsymbol{G}$, $\mathcal{V}$ & $64^3$ & $30 \dots 35$ m & 1000 & \ref{fig:3dcomparison}, \ref{fig:flexible_descriptor} \\
            $\boldsymbol{S}$, $\boldsymbol{G}$, $\mathcal{V}$ & $64^3$ & $45 \dots 55$ m & 1000 & - \\
            $\boldsymbol{S}$, $\boldsymbol{G}$, $\mathcal{V}$ & $128^3$ & $1.5 \dots 2.5$ h & $1000 \dots 1500$ & \ref{fig:fancyimgs} \\
            $\boldsymbol{S}$, $\mathcal{V}$ & $512^3$ & $\approx 3$ d & 1500 & \ref{fig:large} \\
            \bottomrule
        \end{tabular*}
    \end{table}
\end{center}

\subsection{Range of applicability}\label{ssec:results_further-results}
\begin{figure*}[t]
	\centering
	\subfloat[Alloy]{\includegraphics[width=0.24\textwidth]{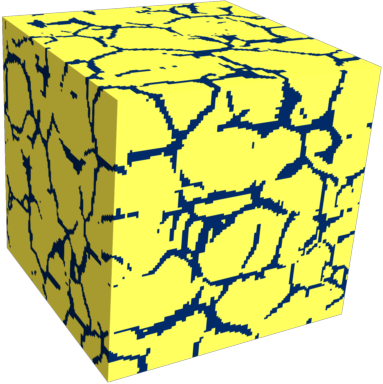}}
	\hfill
	\subfloat[Carbonate]{\includegraphics[width=0.24\textwidth]{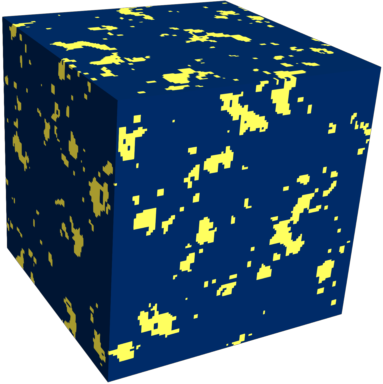}}
	\hfill
	\subfloat[Ceramics]{\includegraphics[width=0.24\textwidth]{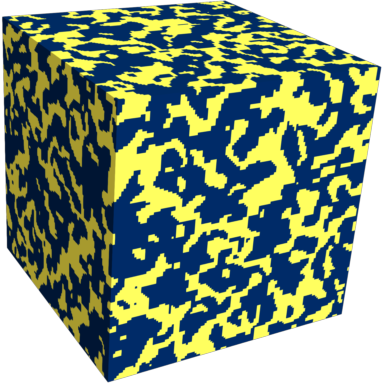}}
	\hfill
	\subfloat[Copolymer]{\includegraphics[width=0.24\textwidth]{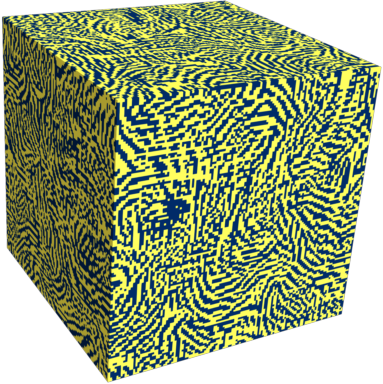}}
	\vfill
	\subfloat[PMMA]{\includegraphics[width=0.24\textwidth]{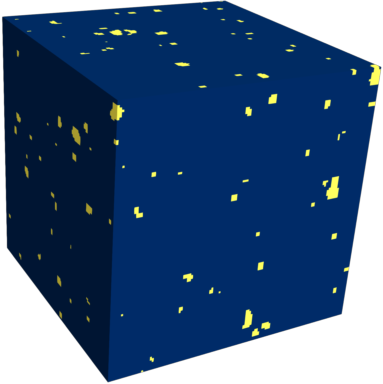}}
	\hfill
	\subfloat[Sandstone]{\includegraphics[width=0.24\textwidth]{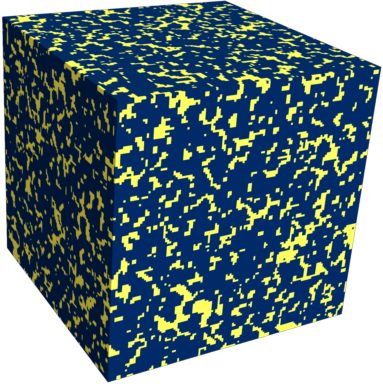}}
	\hfill
	\subfloat[Synthetic]{\includegraphics[width=0.24\textwidth]{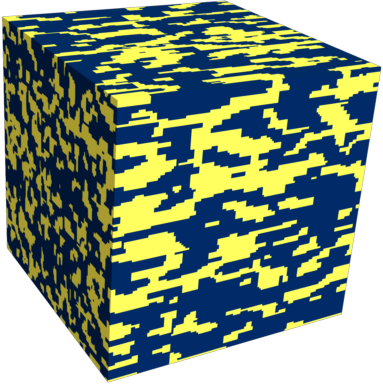}}
	\hfill
	\subfloat[Rubber composite]{\includegraphics[width=0.24\textwidth]{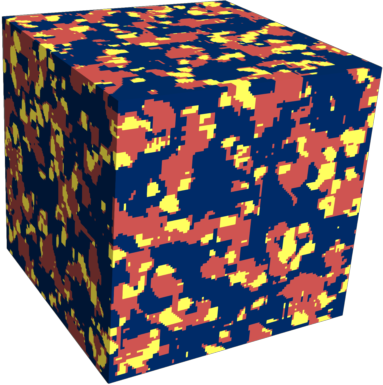}}
	\caption{Various different microstructures reconstructed with \emph{differentiable MCR} using combined descriptors. Note that examples g and h are anisotropic and with three phases respectively. The original structures stem from~\emph{pyMKS}~\cite{brough_materials_2017} and~\cite{li_transfer_2018}. \label{fig:fancyimgs}}
\end{figure*}

A wide variety of different reconstructed microstructures, including anisotropic and three-phase materials, is shown in \autoref{fig:fancyimgs}. For these examples, the descriptor contains spatial two- and three-point correlations, Gram matrices and the total variation, where all values are unrolled into a single large descriptor vector like in \autoref{eq:concatenation}. 
The relative descriptor errors
\begin{equation}
    \mathcal{E} (\boldsymbol{D}(\boldsymbol{M})) = \sum_s^{n^\text{slices}} \dfrac{|| \boldsymbol{D}_s(\boldsymbol{M}) - \boldsymbol{D}^\text{des} ||}{n^\text{slices}|| \boldsymbol{D}^\text{des} ||}
\label{eq:error}
\end{equation}
are given in \autoref{tab:errors}.
Most structures have a very small error, with the PMMA reaching~$\mathcal{E} < 0.01 \%$, but some exceptions (see example a) might require post-processing or better-suited descriptors, as discussed in \autoref{ssec:results_descriptorchoice}.
Generally, the relative correlation error is larger than the relative Gram matrix error. 
In the anisotropic case,~$\mathcal{E}(\boldsymbol{S})$ is lowest in the plane normal to the grain elongation direction, whereas~$\mathcal{E}(\boldsymbol{G})$ is smaller in the other planes.
Overall, the great variety of microstructures and the low errors demonstrate the potential and the range of applicability of the algorithm.
\begin{center}
    \begin{table}[t]%
        \caption{Relative descriptor errors for the reconstructions in \autoref{fig:fancyimgs} as defined in \autoref{eq:error}.\label{tab:errors}}
        \centering
        \begin{tabular*}{\linewidth}{@{\extracolsep\fill}lcc@{\extracolsep\fill}}
            \toprule
            Microstructure & $\mathcal{E}(\boldsymbol{S})$ & $\mathcal{E}(\boldsymbol{G})$  \\
            \midrule
            Alloy & $4.6 \%$ & $0.052 \%$ \\
            Carbonate & $0.092 \%$ & $0.016 \%$ \\
            Ceramics & $0.18 \%$ & $0.013 \%$ \\
            Copolymer & $0.33 \%$ & $0.038 \%$ \\
            PMMA & $0.0025 \%$ & $0.0068 \%$ \\
            Sandstone & $0.022 \%$ & $0.0062 \%$ \\
            Synthetic transversal & $0.43 \%$ & $0.14 \%$ \\
            Synthetic longitudinal & $0.88 \%$ & $0.070 \%$ \\
            Rubber composite & $0.77 \%$ & $0.0046 \%$ \\
            \bottomrule
        \end{tabular*}
    \end{table}
\end{center}

\section{Conclusions}\label{sec:conclusions}
A general framework for reconstructing 3D microstructures from generic prescribed high-dimensional descriptors is presented.
For this purpose, the difference between the current and desired descriptor is defined as a loss to minimize using a gradient-based optimizer.
The formulation does not rely on a training data set of microstructures or on a specific microstructure descriptor. 
Instead, by allowing \emph{any} differentiable descriptor, the generality of the Yeong-Torquato algorithm is coupled with the efficiency of gradient-based optimization.
Three recent, very potent approaches~\cite{li_transfer_2018, bostanabad_reconstruction_2020, bhaduri_efficient_2021} are replicated as special cases.

The presented 3D-formulation satisfies the experimental constraint that real microstructure examples in the form of microscopy images are commonly available on 2D sections only.
Reconstructing 3D microstructures from descriptors of 2D slices leads to noise, which is handled in a new and superior way by reinterpreting a denoising strategy as a physical microstructure descriptor and adapting the formulation accordingly.
This leads to reduced noise, higher robustness with respect to hyperparameters and better reconstruction results.
A series of numerical experiments is performed to explore the potential of the method.
These experiments show the benefit of the proposed denoising strategy and the range of applicability of the method, which is widened by the flexible choice of descriptor.
Although the efficiency of the suggested approach is not on par with random field- and GAN-based approaches, it exceeds that of equally general algorithms by far and lies in a similar order of magnitude as the numerical simulation.
Within the class of general descriptor-based reconstruction algorithms, e.g. the standard Yeong-Torquato algorithm, this substantial improvement overcomes the major hurdle of prohibitively high computational costs.
This major advancement takes the reconstructable image resolution to a new level, now passing the threshold for resolving microstructures in the required size and quality for more representative numerical simulation studies, enabling practical applicability.

Overall, the performance and accuracy of the method is very satisfactory. Due to the generality and simplicity of the formulation, it can serve as a basis for further extensions to achieve a broader range applicability.
Specifically, the following aspects require further attention:
\begin{itemize}
    \item For metallic materials, the lattice orientation information needs to be incorporated.
    \item Additional differentiable descriptors need to be formulated and best-practices of when to apply which descriptor need to be developed.
    \item The authors have found that reconstructions from Gram matrices usually require less iterations than from spatial correlations. The influence of the descriptor formulation on the condition number deserves further attention.
\end{itemize}
With the present formulation and the envisioned extensions, the algorithm allows to generate and visualize 3D microstructures from given 2D micrographs. 
The descriptor-based nature of the algorithm enables to create large data bases of synthetic 3D microstructure from few observations through interpolation in descriptor space. 
This exploration of feasible microstructures leads to valuable insight and computationally-obtained process-structure-property linkages like~\cite{rassloff_accessing_2021} can be established and advanced in order to accelerate materials engineering.


\section*{Acknowledgements}
This research is funded by the European Regional Development Fund (ERDF) and co-financed by tax funds based on the budget approved by the members of the Saxon State Parliament under Grants 100373334.

The authors are grateful to the Center for Information Services and High Performance Computing (Zentrum für Informationsdienste und Hochleistungsrechnen (ZIH)) TU Dresden for providing its facilities for high throughput calculations.

\subsection*{Declaration of competing interest}
The authors declare that they have no known competing financial interests or personal relationships that could have appeared to influence the work reported in this paper.

\subsection*{Data availability}
All data used in this work is freely available and can be obtained from the cited sources.

\subsection*{Code availability}
The code will be made available by the authors upon reasonable request.


\subsection*{CRediT authorship contribution statement}
The development of concepts and algorithms are carried out by P.S. and A.R. under the coordination of M.A. and M.K. The algorithm implementation and numerical computations are performed by P.S. All authors involved in the discussions, the analysis of simulation results and the writing of the manuscript.

\appendix
\section{On pixel-based approximation of smooth surfaces}\label{sec:piis4}
In this appendix, it is discussed why the desired total variation~$\mathcal{V}^\text{des}$, which has been identified as the discretized phase boundary integral in \autoref{ssec:tv_we}, should not be computed from a smooth or tessellated microstructure representation, but rather from a voxel representation.

\begin{figure*}[t]
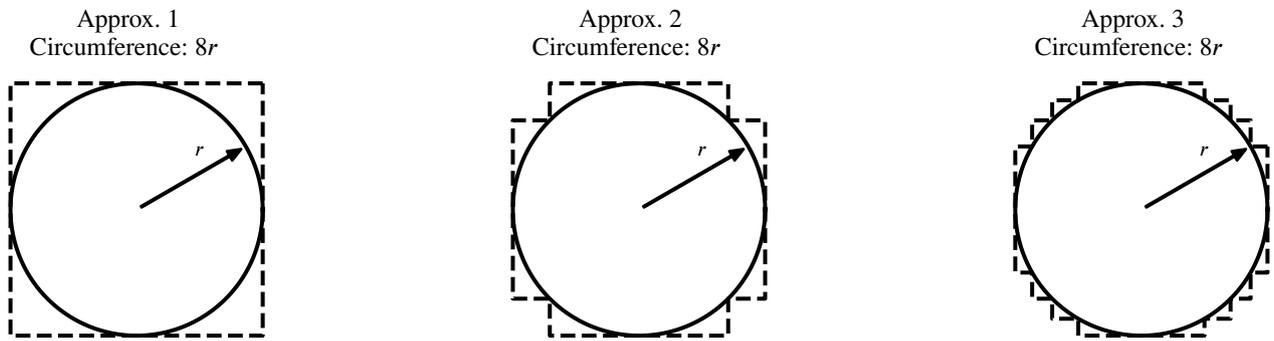

	\centering
	\import{Figures/}{pi_is_4_lvl_1.pgf}
	\hfill
	\import{Figures/}{pi_is_4_lvl_2.pgf}
	\hfill
	\import{Figures/}{pi_is_4_lvl_3.pgf}
	\caption[Pixel-based approximation of smooth curve]{Illustration showing why the desired value of the total variation should stem from a pixel or voxel geometry: Iteratively refining a rectangular approximation to a circle of radius~$r$ does not make the circumference of the approximation converge to~$2\pi r$.\label{fig:piis4}}
\end{figure*}

Consider a circular shape which is approximated through vertical and horizontal lines as shown in Figure~\ref{fig:piis4}. As the approximation becomes increasingly refined (left to right), the shape looks closer to the circle it approximates, and the approximated surface area converges to that of a circle. This is, however, not the case for the circumference, which remains constant irrespective of how much the approximation is refined. This shows that there is a fundamental difference between the Euclidean length of a smooth curve and that of a pixel-based approximation: The latter results in Manhattan distances, whereas the former yields the more common Euclidean distance.
Having identified~$\mathcal{V}$ as a phase boundary integral in Section~\ref{ssec:tv_we}, one might be tempted to directly integrating the phase boundary of a smooth or triangulated microstructure to obtain~$\mathcal{V}^\text{des}$. However, the present considerations illustrate that, unless the Manhattan distance is considered explicitly, the total variation should be computed on a pixel/voxel discretization if it is to be used in reconstruction.

\section{Implementational details}
\label{sec:methods}
The numerical experiments are computed on an~\emph{Nvidia A100} GPU with~\emph{TensorFloat32} disabled and using a single work unit only. 
\emph{TensorFlow}~\cite{abadi_tensorflow_2016} is used for just-in-time compilation in order to generate efficient GPU code from a simple Python definition of the descriptors and the loss function.
Furthermore, the automatic differentiaton in~{TensorFlow} provides the gradient of the loss function with respect to the microstructure.
This gradient is passed to \emph{scipy}'s~\emph{L-BFGS-B}-optimizer~\cite{byrd_stochastic_2015} in order to update the microstructure.
For numerical reasons, all numerical experiments are performed in double precision.

\bibliographystyle{cas-model2-names}
\bibliography{Seibert2021_DMCR-3D.bib}

\end{document}